\documentclass[seceq]{ptptex}

\usepackage{graphicx}
\usepackage{wrapft}
\newcommand{\VEV}[1]{\langle{#1}\rangle}
\newcommand{\chibar}{{\bar{\chi}}}
\newcommand{\Psfig}[3]{\includegraphics[width=#1 #3]{Figs/#2}}
\newcommand{\Feff}{{\cal F}_\mathrm{eff}}
\newcommand{\Mc}{{\cal M}}
\title{%        %You can use \\ for explicit line-break
Strong coupling limit/region of lattice QCD
}

\author{%       %Use \scshape  for the family name
A. \textsc{Ohnishi},
N. \textsc{Kawamoto},
K. \textsc{Miura},
K. \textsc{Tsubakihara},
H. \textsc{Maekawa}
}

\inst{%         %Affiliation, neglected when [addenda] or [errata]
Department of Physics, Faculty of Science, Hokkaido University,\\
Sapporo 060-0810, Japan
}

\abst{%         %this abstract is neglected when [addenda] or [errata]
We study the phase diagram of quark matter and nuclear properties
based on the strong coupling expansion of lattice QCD.
Both of baryon and finite coupling correction are found to have
effects to extend the hadron phase to a larger $\mu$ direction
relative to $T_c$.
In a chiral RMF model with logarithmic sigma potential
derived in the strong coupling limit of lattice QCD,
we can avoid the chiral collapse 
and normal and hypernuclei properties are well described.
}

\begin{document}

\maketitle

\section{Introduction}
\label{Sec:Introduction}

Understanding the properties of nuclei and nuclear matter
from QCD is one of the ultimate goals in nuclear physics.
In a standard roadmap,
it is necessary
to describe hadrons in QCD,
to derive the bear nucleon-nucleon interaction,
to obtain the effective nuclear force,
and to solve nuclear many-body problems.
An alternative way would be
to obtain the effective potential (free energy density) in QCD,
to represent this effective potential as the density functional
in hadronic degrees of freedom,
and to apply this density functional to nuclear many-body problems.
For this purpose, 
the most instructive approach may be
to combine the strong coupling limit (SCL) of lattice 
QCD~\cite{KS1981,SCL-FiniteT,Bilic,KMOO_2007}
and the relativistic mean field (RMF) models,
since Monte-Carlo simulations of lattice QCD for dense matter~\cite{MC_LQCD}
are not yet easy at present.
%
%In fact, effective potential at finite $T$ and $\mu$
%have been analytically derived in SCL,
%and it is successfully applied to nuclear many-body problems.\cite{TO_2007}

In this proceedings,
we discuss the baryon and finite coupling effects on the phase diagram
in the strong coupling region of lattice QCD,\cite{KMOO_2007,QM06-Ohnishi}
and nuclear properties in a chiral RMF model
based on the SCL effective potential.\cite{TO_2007,HYP06-Tsubaki}
%In this proceedings,
%first we derive an expression of the effective potential
%including baryon effects~\cite{KMOO_2007}
%and finite coupling effects up to $1/g^2$
%in the strong coupling limit/region of lattice QCD.
%We find that these effects significantly modifies the shape of
%the phase diagram.
%Next we show some results of the chiral RMF models
%based on the SCL effective potential.\cite{TO_2007,HYP06-Tsubaki}

\section{Effective potential at strong coupling of lattice QCD}
\label{Sec:Model}

The strong coupling limit of lattice QCD (SCL-LQCD) predicts
the high $T$ second order chiral phase transition at $\mu=0$
% in the chiral limit
and the high density first order transition at $T=0$,
and it well explains hadron masses.
While these predictions explain the real world qualitatively,
quantitative understanding of the phase diagram is not achieved yet.
For example, the ratio $R_{\mu T}=\mu_c(T=0)/T_c(\mu=0)$
should be larger than two, but it is less than 1/3
in SCL-LQCD without baryon effects.
In this work, we discuss the effects of baryons~\cite{KMOO_2007}
and finite coupling on the shape of the phase boundary.
%including baryon effects~\cite{KMOO_2007}
%and finite coupling effects up to $1/g^2$

At strong coupling ($g \gg 1$),
%the pure gluonic action ($\propto 1/g^2$) can be treated perturbatively
%in lattice QCD,
the plaquett contribution ($\propto 1/g^2$) is perturbative
and the effective action is obtained
by integrating spatial links as,\cite{SCL-FiniteT}
\begin{align}
S   &= S_\textrm{SCL} + \Delta S_g
	+ \mathcal{O}(1/d, 1/g^2\sqrt{d}, 1/g^4)\ , \\
S_\textrm{SCL} &=\frac12\sum_x \left(e^{\mu}V_x-e^{-\mu}V^\dagger_x\right)
	-\frac12 (M, V_M M)
	-(\bar{B},V_B B)
	+m_0\sum_x M_x\ ,
\label{Eq:ActionA}
	\\
\Delta S_g &=\frac{\beta_t}{2d}
		\sum_{x,j>0} 
		(
		 V^\dagger_x V_{x+\hat{j}}
		+V^\dagger_x V_{x-\hat{j}}
		)
	-\frac{\beta_s}{d-1}
		\sum_{x,k>j>0} 
		M_{x}
		M_{x+\hat{j}}
		M_{x+\hat{k}}
		M_{x+\hat{k}+\hat{j}}
	\ ,
\label{Eq:ActionG}
\end{align}
where
$(A,B)=\sum_x A_xB_x$,
$M_x=\chibar^a_x\chi^a_x$, 
$B_x=\varepsilon_{abc}\chi^a_x\chi^b_x\chi^c_x/6$,
$V_x=\chibar_x U_0(x) \chi_{x+\hat{0}}$,
$\beta_t=d/2N_c^2g^2$,
$\beta_s=(d-1)/8N_c^4g^2$,
$d=3$ is the spatial dimension,
and
$V_M(x,y)$ and $V_B(x,y)$ represent 
mesonic and baryonic inverse propagators.
In SCL,
the effective potential for $N_c=3$ without baryon effects has been known
as,\cite{SCL-FiniteT,Bilic}
\begin{eqnarray}
\Feff^{(T)}(\sigma)
&=&\frac12a_\sigma\sigma^2
+\Feff^{(q)}(a_\sigma\sigma+m_0;T,\mu)
\,,
\label{Eq:FeffT}
\\
\Feff^{(q)}(m_q)
&=&-T\log\left[C_\sigma^3-C_\sigma/2
+\cosh\left(3\mu/T\right)/4\right]
\,,
\end{eqnarray}
where $C_\sigma=\cosh(\mathrm{arcsinh}\,(m_q)/T)$
%where $E_q=\mathrm{arcsinh}\,m_q$ 
%and $a_\sigma=\VEV{V_M}=d/2N_c=1/2$,
and $a_\sigma=d/2N_c$,
and the scalar field $\sigma = - \VEV{M}$
% plays the role of 
is 
the chiral order parameter.
This effective potential
% (\ref{Eq:FeffT})
gives the ratio of $R_{\mu T} \sim 0.33$.

\begin{figure}[b]
\begin{minipage}{\textwidth}
%\Psfig{7cm}{Fig1a.eps}{}
\centerline{
\Psfig{7cm}{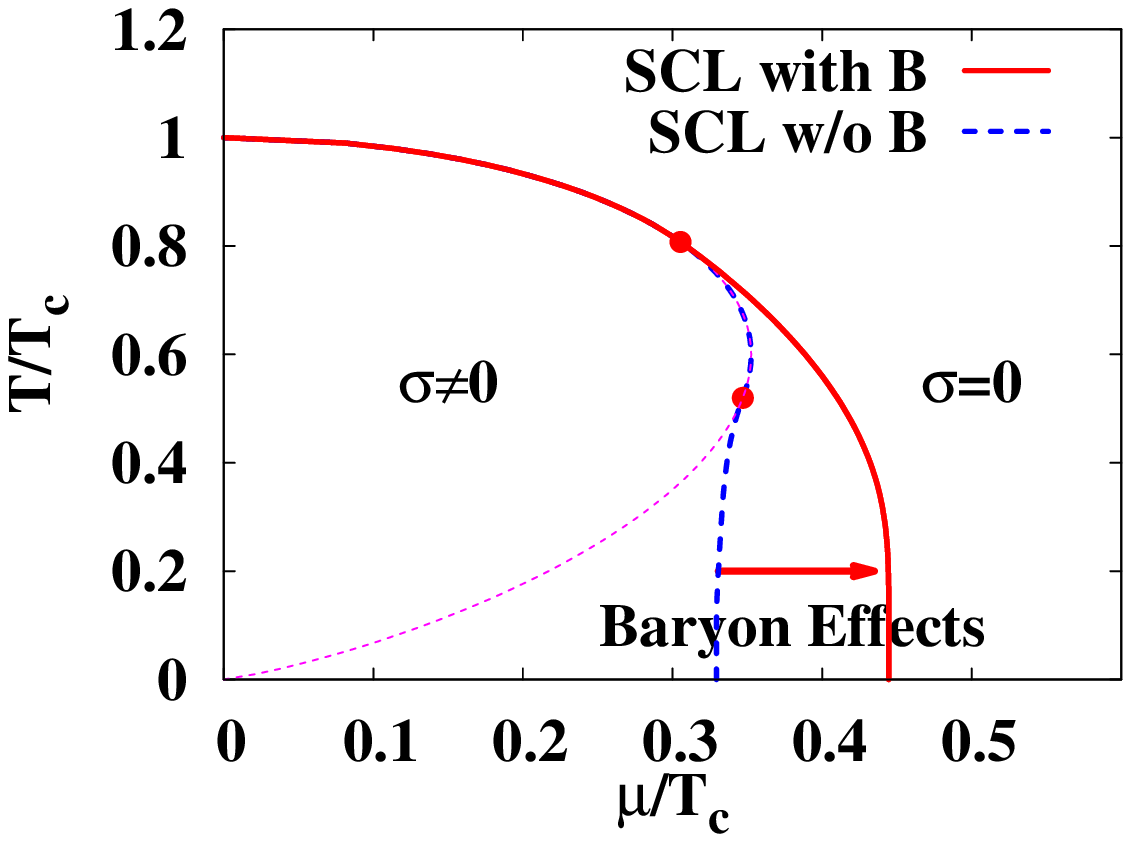}{}
\Psfig{5cm}{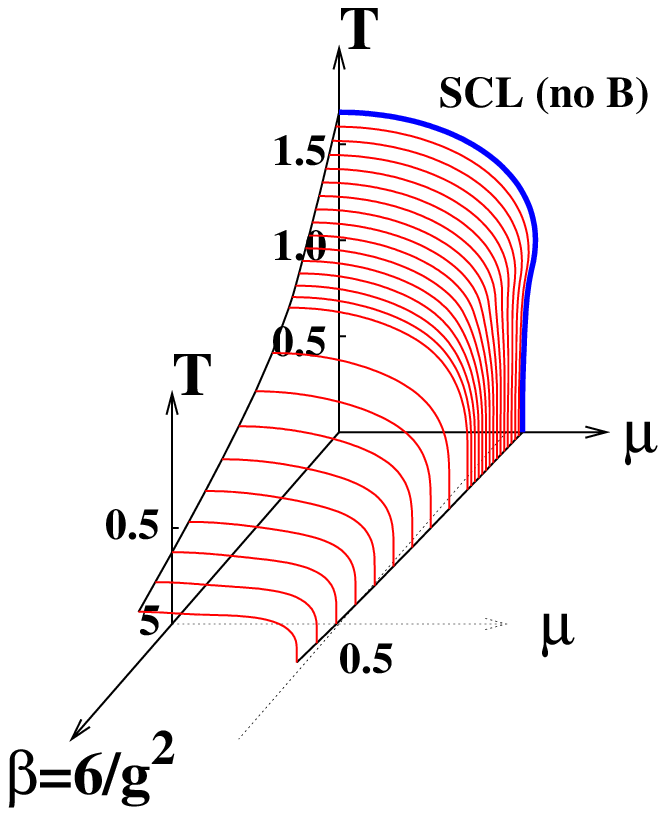}{}
}
\end{minipage}
\caption{
The phase diagram in the strong coupling limit (left panel),
and its evolution with $\beta=6/g^2$ (right panel).
}\label{Fig:SCL}
\end{figure}

%The baryonic composite action
%%, $-(\bar{B},V_B B)$,
%is often ignored 
%%for $N_c \geq 3$,
%%since it is propotional to $1/\sqrt{d^{N_c-2}}$
%in the $1/d$ expansion ($d=3$ is the spatial dimension),
%but 
%%In discussing dense matter, however, 
%baryon effects would be larger in dense matter.
We have recently demonstrated that 
the baryonic composite action can be represented
%in the mesonic composites
in the mean field approximation at zero diquark condensate as,\cite{KMOO_2007}
\begin{eqnarray}
&&e^{(\bar{B},V_B,B)}
\simeq
e^{
	-N_s^3N_\tau\left[a_\omega\omega^2/2
	+\Delta \mathcal{F}^{(b)}_\mathrm{eff}(g_\omega\omega)\right]
	-\sum_x \left[(\alpha^2+\gamma^2)M^2/2+\alpha\omega M)\right]
	}\ ,
\label{Eq:B}
\end{eqnarray}
where auxiliary baryon determinant is represented in 
${\it \Delta}\Feff^{(b)}$.\cite{KMOO_2007}
%${\it \Delta}\Feff^{(b)}(m_b)\simeq-f^{(b)}(\pi m_b/8)$, 
%$f^{(b)}(x)=\log(1+x^2)/2-[\arctan{x}-x+x^3/3]/x^3-3x^2/10$.
At equilibrium,
the baryon potential field $\omega$ is approximately proportional to $\sigma$,
and the effective potential is found to be
\begin{eqnarray}
\Feff^{(Tb)}(\sigma)
&=&\frac12b_\sigma\sigma^2
+{\cal F}_\mathrm{eff}^{(q)}(b_\sigma\sigma)
+{\it \Delta}{\cal F}_\mathrm{eff}^{(b)}(g_\sigma\sigma)
\,.
\label{Eq:Feff}
\end{eqnarray}
Two parameters in this effective potential, $b_\sigma$ and $g_\sigma$,
are related to the decomposition parameters,
$\alpha$ and $\gamma$.
introduced in baryonic composite decomposition.
%\cite{Azcoiti2003}

% Finite coupling correction --------------------------------------------------*

The effective potential with $1/g^2$ correction
was derived by Bili{\'c} {\it et al}.\cite{Bilic},
where $V^\dagger V$ term was generated by the derivative
of $\Feff^{(q)}$.
Since ${\it \Delta} S_g$ is not in the bilinear form
in $\chi$ and $\bar{\chi}$, 
we here bosonize the plaquett contributions
and apply the mean field approximation,\cite{QM06-Ohnishi}
\begin{eqnarray}
{\it \Delta}S_F
&\simeq&
	N_s^3N_\tau\left[
		 \frac{\beta_t}{4}\varphi_t^2
		+\frac{\beta_sd}{4}\varphi_s^2
	\right]
	+\frac{\beta_t\varphi_t}{4}
	\sum_x(V_x-V_x^\dagger)
	-\beta_s\varphi_s
	\sum_{x,j>0} M_x M_{x+\hat{j}}
	\,.
\label{Eq:ActionC}
\end{eqnarray}
The auxiliary fields have expectation values of
$\VEV{\varphi_t}=\VEV{V^\dagger-V}$
and 
$\VEV{\varphi_s}=2\VEV{M_xM_{x+\hat{j}}}$.
These correction terms have a similar structure
to the SCL effective action (\ref{Eq:ActionA}),
and they lead to
the modifications of the quark mass and effective chemical potential as 
$\widetilde{m}_q=\sigma d(1+4N_c\beta_s\varphi_s-\beta_t\varphi_t\cosh\mu)/2N_c$
and
$\widetilde{\mu}=\mu-\beta_t\varphi_t\sinh\mu$.
At equilibrium, 
we can put $\varphi_s=2\sigma^2 + {\cal O}(1/g^2)$,
and the effective free energy up to ${\cal O}(1/g^2)$ without baryon effects
is obtained as,
\begin{eqnarray}
\Feff^{(1/g^2)}=\frac{d}{4N_c}\sigma^2+3d\beta_s\sigma^4
	+\frac{\beta_t}{4}\varphi_t^2-N_c\beta_t\varphi_t\cosh\mu
	+{\cal F}_\mathrm{eff}^{(q)}(\widetilde{m}_q;T,\widetilde{\mu})
	\,.
	\label{Eq:FeffCorr}
\end{eqnarray}

As shown in Fig.~\ref{Fig:SCL}, 
both of baryons and finite coupling corrections
have effects to extend the hadronic phase in the larger $\mu$ direction
relative to $T_c$,
while each of these is not enough to explain the empirical ratio
of $R_{\mu T}$.
It would be interesting to evaluate both of these effects simultaneously.

\section{Chiral RMF with SCL Effective Potential}

%%%%%%%%%%%%%%%%%%%
\begin{wrapfigure}[18]{r}[0pt]{5.8cm}
\centerline{~\Psfig{5.0cm}{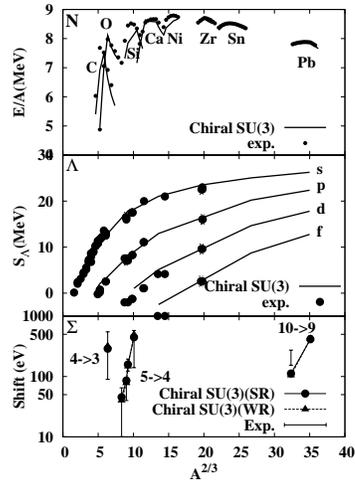}{}~}
\caption{
Nuclear binding energies (top), 
single $\Lambda$ hypernuclear separation energies (middle),
and atomic shift of $\Sigma^-$ atom (bottom)
in the present chiral SU(3) RMF.
}\label{Fig:ChiralRMF}
\end{wrapfigure}
%%%%%%%%%%%%%%%%%%%
RMF models are powerful tools in describing nuclear matter and finite nuclei,
but we have a so-called {\em chiral collapse} problem~\cite{Collapse}
in a naive chiral RMF model based on the $\phi^4$ theory;
the normal vacuum jumps to an abnormal one below the normal nuclear density.
The effective potential in SCL-LQCD gives us a hint to solve this problem.
In a zero temperature treatment of SCL-LQCD,\cite{KS1981,KMOO_2007}
the effective potential is found to have the form,
$\Feff=b_\sigma\sigma^2/2-N_c\log\sigma$,
and the divergent behavior at $\sigma \to 0$ helps to avoid the chiral collapse.

We have recently developed a chiral SU(2) RMF model~\cite{TO_2007}
with logarithmic sigma potential,
\begin{eqnarray}
U_\sigma
&=&
\frac{b_\sigma}{2}\mathrm{tr}(\Mc\Mc^\dagger)-a_\sigma\log\det(\Mc\Mc^\dagger)-c_\sigma\sigma
\nonumber\\
&\sim&
\frac{b_\sigma}{2}\sigma^2-2a_\sigma\log\sigma-c_\sigma\sigma\ ,
\end{eqnarray}
where $\Mc$ denotes the meson matrix.
In this RMF,
we can well describe
symmetric nuclear matter equation of state (EOS)
and bulk properties of finite nuclei.\cite{TO_2007}
In a chiral SU(3) RMF model,\cite{HYP06-Tsubaki}
we include the determinant interaction 
($\det\Mc + \det\Mc^\dagger$)
simulating $\mathrm{U}_A(1)$ anomaly.
After fitting binding energies and charge rms radii of normal nuclei,
we find that the symmetric matter EOS becomes softer
than that in the chiral SU(2) RMF
due to the scalar meson with hidden strangeness, $\zeta=\bar{s}s$,
which couples with $\sigma$ through the determinant interaction.

In this chiral SU(3) RMF,
we can find hyperon-meson coupling constants
which fits existing data of
%binding energies and charge rms radii of normal nuclei,
$\Lambda$ separation energies in single hypernuclei,
$\Lambda\Lambda$ bond energy ($\Delta B_{\Lambda\Lambda}$)
in $^6_{\Lambda\Lambda}$He,
and atomic shifts of $\Sigma^-$ atoms,
as shown in Fig.~\ref{Fig:ChiralRMF}.

While the above sigma potential has divergence at $\sigma \to 0$,
the effective potential derived in the finite $T$ treatments
have a linear term for small values of $\sigma$ at $T \to 0$,
\begin{equation}
\Feff^{(T)} \to \frac{a_\sigma}{2}\sigma^2-N_c\,\mathrm{arcsinh}\,(a_\sigma\sigma)
\ ,
\end{equation}
which is enough to stabilize the normal vacuum.
We may conclude that gluons play decisive roles
to avoid the chiral collapse as pointed out
in Ref.~\citen{glueball}.

\section{Summary}

In this paper,
we have investigated the phase diagram of quark matter
and nuclear properties
based on the strong coupling expansion of lattice QCD.
%and nuclear properties in a chiral SU(3) RMF model
%with the effective potential derived in the strong coupling limit
%of lattice QCD.
%
In the first part, we find that baryons and finite coupling
corrections have favorable effects to extend the hadron phase
to a larger $\mu$ direction with respect to $T_c$.
In the second part, we have shown that 
we can well describe normal and hypernuclear properties
with a logarithmic potential.
For the understanding of dense matter,
it is important to respect
both of chiral symmetry and strangeness degrees of freedom.
In addition, the present work may be suggesting the importance
of implicit role of gluons, which generate the effective potential
of hadrons.

%\section*{Acknowledgements}
This work is supported in part by the Ministry of Education,
Science, Sports and Culture,
Grant-in-Aid for Scientific Research
under the grant numbers,
    13135201,		% Kawamoto
    15540243,		% ((C)(2), 2003), Ohnishi
and 1707005.		% (Tokutei, Strangeness)

%\appendix
%\section{} %Empty argument \section{} yields `Appendix'. 

%%%%%%%%%%%%%%%%%%%%%%%%%%%%%%%%%%%%%%%%%%%%%%%%%%%%%%%%%%%%%
% Some macros are available for the bibliography:
%  o for general use
%    \JL : general journals                 \andvol : Vol (Year) Page
%  o for individual journal 
%    \AJ   : Astrophys. J.           \NC         : Nuovo Cim.
%    \ANN  : Ann. of Phys.           \NPA, \NPB  : Nucl. Phys. [A,B]
%    \CMP  : Commun. Math. Phys.     \PLA, \PLB  : Phys. Lett. [A,B]
%    \IJMP : Int. J. Mod. Phys.      \PRA - \PRE : Phys. Rev. [A-E]     
%    \JHEP : J. High Energy Phys.    \PRL        : Phys. Rev. Lett.
%    \JMP  : J. Math. Phys.          \PRP        : Phys. Rep.
%    \JP   : J. of Phys.             \PTP        : Prog. Theor. Phys.     
%    \JPSJ : J. Phys. Soc. Jpn.      \PTPS       : Prog. Theor. Phys. Suppl.
% Usage:
%  \PRD{45,1990,345}          ==> Phys.~Rev.\ \textbf{D45} (1990), 345
%  \JL{Nature,418,2002,123}   ==> Nature \textbf{418} (2002), 123
%  \andvol{B123,1995,1020}    ==> \textbf{B123} (1995), 1020
%%%%%%%%%%%%%%%%%%%%%%%%%%%%%%%%%%%%%%%%%%%%%%%%%%%%%%%%%%%%%

\end{document}